**SUBMITTED TO NMR IN BIOMEDICINE**

**RADICAL-FREE HYPERPOLARIZED MRI USING ENDOGENOUSLY-OCCURING PYRUVATE ANALOGUES AND UV-INDUCED NONPERSISTENT RADICALS**

Claudia C Zanella[1], Andrea Capozzi[1], Hikari AI Yoshihara[1], Alice Radaelli[1], Lionel P Arn[2], Rolf Gruetter[1], Jessica AM Bastiaansen[2]

[1] Laboratory of Functional and Metabolic Imaging, EPFL, Lausanne, Switzerland.

[2] Department of Diagnostic and Interventional Radiology, Lausanne University Hospital (CHUV) and University of Lausanne (UNIL), Lausanne, Switzerland

**To whom correspondence should be addressed:** Jessica AM Bastiaansen, Department of Diagnostic and Interventional Radiology, University Hospital Lausanne (CHUV) and University of Lausanne (UNIL), Rue de Bugnon 46, BH 08.74, 1011 Lausanne, Switzerland, Phone: +41-21-3147516, Email: jbastiaansen.mri@gmail.com, Twitter: @jessica_b_

**Sponsors and grant numbers**: JB received funding from the Swiss National Science Foundation (grant number PZ00P3_167871), the Emma Muschamp foundation, and the Swiss Heart foundation (grant number FF18054).

**List of Abbreviations:**

| | |
|---|---|
| $\alpha$kB | alpha-ketobutyric acid (2-oxobutanoic acid) |
| $\alpha$kV | alpha-ketovaleric acid (2-oxopentanoic acid) |
| BA | butyric acid (butanoic acid) |
| dDNP | dissolution dynamic nuclear polarization |
| ESR | electron spin resonance |
| FM | frequency modulation |
| Glc | glucose |
| HP | hyperpolarized |
| PA | pyruvic acid (2-oxopropanoic acid) |
| SNR | signal-to-noise ratio |
| TEMPOL | 1-$\lambda_1$-oxidanyl-2,2,6,6-tetramethylpiperidin-4-ol |
| TR | repetition time |
| UV | ultraviolet |
| UV-Vis | ultraviolet-visible |

**ABSTRACT**

It was recently demonstrated that nonpersistent radicals can be generated in frozen solutions of metabolites such as pyruvate by irradiation with ultraviolet (UV) light, enabling radical-free dissolution DNP. Although pyruvate is endogenous, an excess of additional pyruvate may perturb metabolic processes, making it potentially unsuitable as a polarizing agent when studying fatty acids or carbohydrate metabolism. Therefore, the aim of the study was to characterize solutions containing endogenously-occurring alternatives to pyruvate as UV-induced nonpersistent radical precursors for *in vivo* hyperpolarized MRI.

The metabolites alpha-ketovalerate ($\alpha$kV) and alpha-ketobutyrate ($\alpha$kB) are analogues of pyruvate and were chosen as potential radical precursors. Sample formulations containing $\alpha$kV and $\alpha$kB were studied with UV-visible spectroscopy, irradiated with UV light, and their nonpersistent radical yields were quantified with ESR and compared to pyruvate. The addition of $_{13}$C labeled substrates to the sample matrix altered the radical yield of the precursors. Using αkB increased the $_{13}$C-labeled glucose liquid state polarization to 16.3 $\pm$ 1.3% compared with 13.3 $\pm$ 1.5% obtained with pyruvate, and 8.9 $\pm$ 2.1% with $\alpha$kV. For [1-$_{13}$C]butyric acid, polarization levels of 12.1 $\pm$ 1.1% for $\alpha$kV and 12.9 $\pm$ 1.7% for $\alpha$kB were achieved. Hyperpolarized [1-$_{13}$C]butyrate metabolism in the heart revealed label incorporation into [1-$_{13}$C]acetylcarnitine, [1-$_{13}$C]acetoacetate, [1-$_{13}$C]butyrylcarnitine, [5-$_{13}$C]glutamate and [5-$_{13}$C]citrate.

This study demonstrates the potential of αkV and αkB as endogenous polarizing agents for *in vivo* radical-free hyperpolarized MRI. UV-induced, nonpersistent radicals generated in endogenous metabolites enable high polarization without requiring radical filtration, thus simplifying the quality-control tests in clinical applications.

## INTRODUCTION

Hyperpolarization via dissolution dynamic nuclear polarization (DNP) enables a several orders of magnitude increase of the polarization of nuclear spins[1]. Hyperpolarized $^{13}$C-enriched probes have enabled real-time imaging of metabolic pathways *in vivo*[2–6]. More recently, a successful translation to human subjects was achieved[7], demonstrating the potential of hyperpolarized MR for several clinical applications[8–12].

Hyperpolarization via DNP requires the presence of polarizing agents which provide the electron spin polarization to the nuclear spins via microwave irradiation at the appropriate frequency. Typically, unpaired electrons of persistent radicals are used as polarizing agents, which have proven to be highly efficient for dissolution DNP[13,14]. This poses a major challenge for the translation to clinical applications: free radicals may be toxic for living organisms and require scavenging or mechanical filtration before a time-consuming quality-control step prior to injection[15]. This shortens the brief measurement window of the metabolic imaging experiment, limited by the relaxation of the hyperpolarized state back to thermal equilibrium[3,16]. Employing nonpersistent radicals generated via ultraviolet (UV) light irradiation of particular precursor molecules may address this challenge. Nonpersistent radicals recombine into diamagnetic and biocompatible species at 190 K[17] and are thereby eliminated instantly during the dissolution process, resulting in radical-free hyperpolarized solutions. This obviates the need for filtration of the endogenous radical precursor.

Several candidates have been demonstrated as suitable precursors for UV-induced nonpersistent radicals, namely a mixture of butanol and phenol[18], pyruvic acid (PA)[19–22], [2-$^{13}$C]PA[23], phenylglyoxylic acid[24] and trimethylpyruvic acid[25]. So far, a few *in vivo* studies have demonstrated the use of UV-induced nonpersistent radicals to measure *in vivo* metabolic processes[19,21,24]. While phenylglyoxylic acid was demonstrated to be beneficial as polarizing agent for photosensitive metabolites[24], the achievable polarization levels were relatively low, at least at 3.35 T and 1.25 K. The use of PA as polarizing agent achieved higher polarization levels[19–

[22], but its presence may interfere or even compete with metabolic processes that involve or are linked to pyruvate dehydrogenase activity[6,26–28]. In the heart, measuring the metabolism of fatty acids is important as it constitutes its main energy source[29,30]. Using PA as a polarizing agent in such hyperpolarized MR studies may not be suitable, as demonstrated by the observed change in cardiac metabolism of the hyperpolarized short-chain fatty acid butyrate in the presence of co-injected pyruvate[6]. Therefore, it is of interest to find endogenously-occuring alternatives to PA for their use as nonpersistent radical precursors for the measurement of short- or medium-chain fatty acid metabolism with for example, hyperpolarized acetate[31,3231], butyrate[6,33–36], or octanoate[37]. Additionally, such alternatives may be beneficial for studies where glucose[38], lactate[39,40] or alanine[41] are hyperpolarized and the formation of pyruvate is a metabolic product of interest.

Alpha-ketobutyrate ($\alpha$kB) and alpha-ketovalerate ($\alpha$kV) are two analogues of PA that naturally occur in human blood[42,43]. All three metabolites belong to the group of $\alpha$-keto acids carrying different biochemical properties[34,43], with $\alpha$kB approximately a 50% less efficient substrate of the enzyme pyruvate dehydrogenase (PDH) when compared with PA, while $\alpha$kV is neither a substrate nor inhibitor of the enzyme[43]. Because it has been demonstrated that UV-irradiation of $\alpha$-keto acids creates radicals[25], $\alpha$kB and $\alpha$kV may have a high potential to be used as nonpersistent radical precursors.

The aim of this study was to characterize the endogenous pyruvate analogues $\alpha$kV and $\alpha$kB as nonpersistent radicals following UV irradiation and to quantify their potential as endogenous polarizing agents for dissolution DNP. Effects of matrix composition on radical yields and polarization levels of $_{13}$C labeled butyric acid and glucose were quantified and a comparison was made with PA. In a proof-of-concept *in vivo* study it was investigated whether $\alpha$kV and $\alpha$kB could be used to measure cardiac metabolism of $_{13}$C-labeled BA.

**EXPERIMENTAL**

*Sample formulation and preparation*

All chemicals were ordered from Sigma-Aldrich (Buchs, SG, Switzerland). Different sample formulations were used depending on the type of experiment.

**Ultraviolet-visible spectroscopy (UV-Vis)** experiments were performed at room temperature on 100 mM of $\alpha$kV, $\alpha$kB or PA in glycerol-water for characterizing UV light absorption of the radical precursors.

**Electron spin resonance (ESR)** was used to characterize the ESR line-shape and to quantify the concentration of the photo-induced radicals. ESR experiments were performed with 5 M solutions of $\alpha$kV, $\alpha$kB or PA in glycerol:water (GW) 1:1 (v/v). The ESR signal intensity was calibrated using six glycerol-water solutions with known TEMPOL concentrations between 0 mM and 100 mM (see Supporting Information, Figure S1). In a second series of experiments, 2 M unlabeled glucose (Glc) or 5.7 M butyric acid (BA) was added to glycerol-water and the amount of radical precursor was empirically optimized to obtain 40 mM of nonpersistent radicals after 200 s of UV irradiation, to ensure comparability of DNP experiments (see Supplementary Information for more details on the empirical optimization procedure). Setting the target radical concentration to 40 mM was a choice made based on previous experience with broad line-width radicals used to hyperpolarize $_{13}$C-labled nuclei at 7T[39,44].

**Hyperpolarized $_{13}$C MRS** was performed on samples containing fully deuterated, fully $_{13}$C labeled glucose ([U-$_{13}$C$_6$, U-$_2$H$_7$]Glc) or [1-$_{13}$C]butyric acid ([1-$_{13}$C]BA). After optimization on samples prepared with non-labeled compounds, the amount of radical precursor was set to 5.7 M $\alpha$kV, 4.1 M $\alpha$kB and 1.6 M PA for samples containing [U-$_{13}$C$_6$, U-$_2$H$_7$]Glc. Conversely, the radical precursor was set to 2.4 M $\alpha$kV and 4.0 M $\alpha$kB for samples containing [1-$_{13}$C]BA. The latter were also used for *in vivo* measurements.

*Creating nonpersistent radicals on metabolites using UV irradiation*

To create nonpersistent radicals, sample formulations described in the previous section were sonicated and degassed at 50°C for 20 minutes prior to pipetting 6 $\mu$l droplets and freezing them in liquid nitrogen to create a solid pellet. The frozen beads, were transferred to a quartz Dewar (Magnettech, Freiberg Instruments, Germany) filled with liquid nitrogen and irradiated with a broadband UV-lamp (Dymax BlueWave 200, Torrington, CT, USA) at maximum power (40 Wcm$^{-2}$) for a maximum of 200 s using a home-built irradiation setup[22].

In experiments aiming at investigating the time course of radical generation for $\alpha$kV, $\alpha$kB or PA (*n* = 3, each precursor), frozen beads were irradiated for a set duration (i.e. 20 s, 45 s, 70 s, 110 s, 150 s, 200 s), and ESR was measured at the end of each step.

*Quantification and characterization of nonpersistent radicals*

UV-Vis spectroscopy was performed to measure the light absorbance of the different radical precursors using a single beam UV-3100PC Spectrophotometer (VWR International) and a 1 mm pathlength quartz cuvette. UV light absorption of 100 mM $\alpha$kV, $\alpha$kB or PA in glycerol-water samples was measured from 280 – 600 nm in steps of 0.5. Measurements were performed at room temperature since there was no significant difference between absorbance spectra of PA solutions acquired at room- and liquid-nitrogen temperature[22].

ESR was used to determine the nonpersistent radical concentration generated in the samples after UV irradiation at liquid-nitrogen temperature. X-band ESR was performed at 77 K as well using a MiniScope MS 400 spectrometer (Magnettech GmbH, Germany). Spectrometer parameters were chosen to ensure that saturation of the ESR signal was avoided over the entire range of radical concentrations, and kept constant throughout all experiments. Parameters were set to: 20 s sweep time, 20 mT magnetic field range, 0.2 mT magnetic field modulation amplitude and 30 dB power attenuation. ESR experiments were performed on two 6 $\mu$l beads for each sample formulation. Subsequently, the beads were extracted from the quartz Dewar and transferred to a pre-weighted microcentrifuge tube, which was then weighted to determine their exact volume and correct the concentration calibration.

### *Hyperpolarization via DNP*

All DNP experiments were performed in a 7 T custom-built polarizer. Nuclear spins were hyperpolarized using a millimeter-wave source with digital control for frequency modulation, a 55 mW output power and a 1 GHz kHz tuning range (Elva-1 VCOM-06/197/1.0/55-DD). Experiments were performed to determine optimal hyperpolarization conditions in terms of microwave irradiation frequency and to quantify polarization build-up times using the three photo-induced nonpersistent radicals. The microwave frequency sweeps were performed with and without microwave frequency modulation (FM) to find the optimal microwave frequency and to quantify the effect of microwave FM on the DNP enhancement. The sample cup was filled with 12 frozen UV-irradiated beads containing [U-$_{13}$C, U-$_2$H]Glc. Microwave frequency modulated profiles were acquired at 4.2 K using a constant 40 MHz modulation amplitude and 5 kHz modulation rate for each microwave frequency step. The step size was 40 MHz for beads containing $\alpha$kV or $\alpha$kB, and 20 MHz for beads containing PA. For each frequency step, the sample was irradiated for 40 minutes and the polarization build-up was monitored using hard 2° RF excitation pulses every 5 minutes.

The optimal conditions found in the previous experiments for a given set of frequency modulation parameters were used to hyperpolarize [U-$_{13}$C, U-$_2$H]Glc as well as [1-$_{13}$C]BA for quantification of $_{13}$C polarization levels in liquid state after dissolution and for the *in vivo* experiments. The $_{13}$C-labeled metabolites were hyperpolarized at 1.05 $\pm$ 0.02 K for 2.5 hours, with frequency modulated (40 MHz amplitude at a rate of 5 kHz) microwave irradiation at 196.69

GHz center frequency for $\alpha$kV and $\alpha$kB, and at 196.65 GHz center frequency for PA. Their polarization build-up was monitored using 2° RF excitation pulses every 5 minutes.

### *Dissolution DNP, hyperpolarized $_{13}$C MRS in phantoms and in vivo*

Frozen hyperpolarized beads were dissolved using either 5.5 ml of D$_2$O or a phosphate buffered saline solution for liquid state *in vitro* and *in vivo* experiments respectively[44,45]. The dissolved sample was automatically transferred to a separator/infusion pump located in a 9.4 T horizontal bore magnet (Agilent, Palo Alto, CA, USA)[46]. Hyperpolarized $_{13}$C MR spectra were recorded within the pump using a dual $_1$H/$_{13}$C volume coil starting 3 s after dissolution and using a 5° RF excitation pulse with 3 s repetition time (*TR*). After a complete decay of the hyperpolarized magnetization, a thermal equilibrium $_{13}$C spectrum of the sample was acquired using a 90° RF excitation pulse, with a *TR* of 60 s and 64 averages. The enhancement $\epsilon$ was calculated as the ratio of the hyperpolarized and thermal signal peak integral, taking into account a correction for the RF excitation angle and number of averages. The remaining $_{13}$C hyperpolarization after dissolution and transfer was calculated as $P = \epsilon * \tanh\left(\frac{\hbar \gamma_C B_0}{2 k_B T}\right)$, where the hyperbolic tangent represents the $_{13}$C thermal equilibrium polarization at 293 K and 9.4 T.

***In vivo*:** Hyperpolarized experiments were performed *in vivo* to demonstrate a proof of concept of the use of the novel polarizing agents to measure cardiac metabolism of hyperpolarized [1-$_{13}$C]BA. *In vivo* experiments were conducted on two male Wistar rats according to federal ethical guidelines and were approved by the local regulatory body. Experimental set-up including a description of the isoflurane anesthesia followed the method described previously[6,21]. Animal physiology was monitored throughout the experiment and included tracking cardiac and respiration rhythms via a femoral artery catheter as well as of body temperature, which was maintained at a physiological level using a tubing system with circulating warm water. A volume of 62 µl [1-$_{13}$C]BA was hyperpolarized, a volume similar to previous *in vivo* experiments using [1-

$_{13}$C]BA [6,21], according to the procedure described in the previous section. Frozen droplets of 10 M NaOH solution were added to the sample cup to neutralize the hyperpolarized solution during dissolution. Following an automated dissolution and transfer to the separator infusion pump[46], which was prefilled with 0.6 ml of phosphate buffered saline and heparin, 0.8 ml of the hyperpolarized solution was administered via a femoral vein catheter.

MR data was recorded using a custom-made RF hybrid probe of $_1$H/$_{13}$C-pair surface coils. Correct positioning of the coil on the chest of the animal in supine position was ensured using gradient echo $_1$H MRI. FAST(EST)MAP shimming was performed until a $_1$H linewidth of 30 Hz was achieved. Respiratory gated and cardiac-triggered MR acquisitions were performed using a $_1$H-decoupled (WALTZ-16)[47] sequence with adiabatic 30° RF excitation pulses (BIR-4)[48] and a TR of 3 s. Furthermore, 8258 complex data points were acquired to sample a bandwidth range of 20.5 kHz. The first spectra acquired following injection, in which metabolic products are absent, were used to confirm the resonances identified in phantom experiments (Supplementary material, Figure 2).

### *Data processing and analysis*

The ESR concentration calibration curve was obtained from fitting the second integrals of the ESR signal intensity linearly as a function of the radical concentration (Supplementary material, Figure 1). To calculate the build-up rate of nonpersistent radical formation, the second integral of the ESR signal intensity was corrected for bead volume variation prior to performing an intensity calibration to quantify absolute concentrations.

Following methods previously described[22], the radical generation time course was fit to a mono-exponential function to extract the radical generation rate constant and plateau value.

Regarding the DNP sweeps, the y-axis value at each frequency point corresponded to the polarization plateau measured at 4.2 K. The non-modulated sweeps were normalized to 1 and the relative increase in DNP enhancement due to microwave modulation was then calculated accordingly. The microwave frequencies at which highest polarization levels are achieved are referred to as the negative DNP maximum and the positive DNP maximum.

For each sample, the polarization time course at the best DNP condition was fit to a mono-exponential function in order to extract build up time constant and the polarization plateau, as is customary. Accordingly, each sample was irradiated for at least 3 time constants prior to dissolution.

Statistical significant differences between polarization levels obtained using $\alpha$kB or $\alpha$kV compared with PA were tested via an unpaired, 2-tail t-test assuming equal variance, with $p < 0.05$ considered significant.

*In vivo* spectra were post-processed in VnmrJ 3.2 (Agilent, Palo Alto, CA, USA) using 20 Hz line-broadening, baseline correction and drift correction. The signal-to-noise ratio (SNR) was calculated as the ratio of the highest signal intensity after phasing and the standard deviation of the noise over a region without metabolic or injected substrate resonances. All *in vivo* spectra where the [1-$^{13}$C]acetylcarnitine resonance was visible were summed, corresponding to 10 consecutive spectra in the case of $\alpha$kV, and 12 spectra in the case of $\alpha$kB. Chemical shifts were assigned using [1-$^{13}$C]acetylcarnitine as reference peak resonating at 173.9 ppm and assigning all other metabolites as indicated before[6].

**RESULTS**

To characterize structural changes and absorption characteristics of UV-light irradiation on the metabolites αkV, αkB and PA, UV-Vis and ESR measurements were performed: UV-Vis absorption spectra showed a ~1.7x higher absorbance for αkV compared with αkB or PA (Figure 1a). αkB and PA showed nearly identical absorbance maxima in the UV range between 300 – 400 nm (Figure 1a). Absorbance of all three metabolites peaked around a wavelength of 320 nm.

ESR performed on frozen samples prior to UV irradiation indicated the initial absence of unpaired electron spins in the matrixes of glycerol-water mixed with 5 M of αkV, αkB or PA. ESR spectra acquired after 200 s of UV irradiation demonstrated that free radicals were generated within the frozen samples (Figure 1b-d). ESR spectra of αkV and αkB showed a nearly identical shape (Figure 1b,c) but distinct to the PA spectrum (Figure 1d). The production of nonpersistent radicals as function of UV-irradiation time followed a near mono-exponential build-up (Figure 1e) with a characteristic time constant of $30.9 \pm 5.1$ s for αkV, $37.0 \pm 5.2$ s for αkB and $46.5 \pm 1.4$ s for PA (Table 1). The maximum nonpersistent radical concentrations were $41.6 \pm 0.6$ mM for αkV, $56.1 \pm 2.7$ mM for αkB and $55.0 \pm 1.9$ mM for PA. Irradiating the samples for 200 s resulted in plateauing radical concentration while avoiding pulverization of the beads due to excessive UV-irradiation.

To obtain transparent glassy beads that remained intact upon irradiation with UV-light, 2 M Glc were dissolved with glycerol-water and admixed with either 5.7 M of αkV, 4.1 M of αkB and 1.6 M of PA. These formulations yielded a nonpersistent radical concentration of $41.5 \pm 2.5$ mM in αkV, $39.5 \pm 2.3$ mM αkB and $41.7 \pm 2.0$ mM in PA after 200 s UV irradiation (Figure 2a).

To assess the effect of FM modulation of the microwave irradiation the $^{13}$C nuclear polarization was measured as a function of the microwave frequency, which showed for the αkV and αkB sweeps a broadening when FM modulation of the microwave was applied. Modulation of the microwave frequency increased the DNP performance in terms of signal enhancement of hyperpolarized $_{13}$C labeled glucose by 100%, 50% and 30% for αkV, αkB and PA, respectively (Figure 3, Table 2). For αkV and αkB / PA, the microwave frequencies of the positive and negative DNP maxima were observed at: $\nu_{max}$ = 196.69 GHz / 196.65 GHz, $\nu_{min}$ = 196.89 GHz / 196.89 GHz, respectively (Figure 3, Table 2).

To determine the buildup rates of nuclear magnetization, the latter was measured as a function of microwave irradiation duration. The solid state build-up times of hyperpolarized [U-$_{13}$C, U-$_2$H]Glc were 1.7k ± 0.5k s, 1.3k ± 0.5k s and 1.6k ± 0.5 k s for αkV, αkB and PA, respectively (Table 3)

Following dissolution, the liquid state polarization for the C$_{2-5}$ group of [U-$_{13}$C, U-$_2$H]Glc were 8.9 ± 2.1%, 16.3 ± 1.3% and 13.1 ± 1.5% for αkV, αkB and PA, respectively (Table 3, Figure 4). While [U-$_{13}$C, U-$_2$H]Glc hyperpolarized with αkB showed significantly higher liquid state polarization than those hyperpolarized using PA (p = 0.048), there was no significant difference between PA and $\alpha$kV (p = 0.09). The quantification of polarization levels on the C$_1$ and C$_6$ resonances of glucose did not alter the results (Table 3).

The solid state build-up times of hyperpolarized [1-$_{13}$C]butyrate were 2.1 ± 0.3 ks and 3.3 ± 0.4 ks when hyperpolarized using αkV and αkB, respectively. Liquid state enhancement over thermal polarization at 9.4 T was 14.6 ± 1.4k and 15.6 ± 1.7k for αkV and αkB samples, respectively, which translated to liquid state polarization of 12.1 ± 1.1% and 12.9 ± 1.7% (Table 4). In these experiments the natural abundance $_{13}$C resonances of $\alpha$kV and $\alpha$kB could be identified as [1-$_{13}$C]$\alpha$kV (at 172.9 ppm), [1-$_{13}$C]$\alpha$kV-hydrate (at 180.7 ppm), [2-$_{13}$C]$\alpha$kV (at 209.1 ppm), [1-$_{13}$C]$\alpha$kB (at 171.9 ppm), [1-$_{13}$C]$\alpha$kB-hydrate (at 180.6 ppm) and [2-$_{13}$C]$\alpha$kB (at 208.4 ppm) (Figure 5). Note that pH was not neutralized in these experiments, and the chemical shifts were different in the pH-neutralized *in vivo* experiments (Table 5).

To assess the potential to measure cardiac metabolism hyperpolarized [1-$_{13}$C]butyrate was injected in male Wistar rats. The first few seconds after injection of the hyperpolarized solution, metabolic products were not observable and the detected $_{13}$C resonances could be identified as [1-$_{13}$C]butyrate (at 184.8 ppm), and natural abundance resonances of [1-$_{13}$C]$\alpha$kV (at 172.0 ppm), [2-$_{13}$C]$\alpha$kV (at 208.7 ppm), [1-$_{13}$C]$\alpha$kB-hydrate (at 178.1 ppm), [1-$_{13}$C]$\alpha$kB (at 172.1 ppm) and [2-$_{13}$C]$\alpha$kB (at 209.3 ppm). The resonance of [1-$_{13}$C]$\alpha$kV-hydrate was not detected *in vivo*, further evidenced by its absence immediately after injection (Supplementary materials, Figure S2). Maximum *in vivo* SNR on $_{13}$C BA was observed 12 s after dissolution, with a SNR of 1370 for αkV and 1780 for αkB. Cardiac metabolism of hyperpolarized [1-$_{13}$C]butyric acid (Figure 6, Table 5) resulted in $_{13}$C labeling of [1-$_{13}$C]acetylcarnitine (173.9 ppm), [1-$_{13}$C]acetoacetate (176.0 ppm), [1-$_{13}$C]butyrylcarnitine (176.4 ppm), [5-$_{13}$C]glutamate (182.4 ppm), and [5-$_{13}$C]citrate (179.8 ppm). Due to the shoulder of the nearby $\alpha$kB-hydrate resonance, [5-$_{13}$C]citrate could not be detected in the $\alpha$kB experiment.

**DISCUSSION**

This study shows that $\alpha$kV and $\alpha$kB can be used to generate nonpersistent radicals for hyperpolarizing $_{13}$C-labeled substrates, extending the method of radical-free dissolution DNP performed on mixtures containing only endogenously-occurring substances. Their potential as polarizing agents for *in vivo* metabolic studies was also shown in proof-of-concept experiments performed in the heart.

Although UV-irradiated $\alpha$kV and $\alpha$kB demonstrated virtually identical ESR line-shapes, their reaction to UV-light visibly differed (Figure 1). This was not only seen in the different UV-Vis absorption at 100 mM but also in their different radical yield at 5 M. Adding glucose or butyric acid further changed the reaction of each compound to UV light such that a specific sample formulation was required to achieve the targeted nonpersistent radical concentration of 40 mM. As has been observed before[22], the relation between nonpersistent radical yield and the concentration of the precursor was non-linear. This posed a challenge during sample formulation optimization. Even though $\alpha$kV, $\alpha$kB and PA differ from each other by additional methylene units on their aliphatic side chain, the results from our experiments illustrate that sample formulation requires a careful optimization in terms of UV-induced radical yield and polarization level for each $_{13}$C-labeled metabolic substrate, which is an empirical and nontrivial process. Because the current work focused on the feasibility of $\alpha$kV and $\alpha$kB as nonpersistent radical precursors, optimization of the sample formulation did not go beyond obtaining the targeted concentration of 40 mM nonpersistent radicals using a highly empirical approach, resulting in relatively large concentrations of precursor compared with PA in the $_{13}$C-labeled sample formulations. It is important to point out that samples were not specifically optimized to maximize radical yield, to minimize precursor concentrations or to maximize achievable polarization levels. The empirical approach of optimizing sample formulations, plus the observation that radical quantum yields were comparable in glycerol-water mixtures (Figure 1), show that there is significant room for improvement. Our experience with UV-irradiated nonpersistent radical precursors and published work[21,22] indicate that precursor volumes can be reduced when boosting radical yield or UV-light penetration, which may be achieved by using different glassing agents such as ethanol, by deuteration of the radical precursors, by changing the relative ratios of precursor to glassing agent, by adapting the UV irradiation time or changing the bead size.

The nuclear polarization levels increased when applying microwave frequency modulation, which is a consequence of relatively broad radical ESR spectrum and short electron $T_1$ characterizing the UV-radical family (approximately 100 ms)[49–51]. In contrast, microwave FM has little effect on narrow linewidth radicals and radicals with long electron $T_1$ such as OXO63 at 6.7 T[52]. Polarization levels in the current study were on the order of 9% to 16% for glucose and 16% for butyric acid. This was achieved with mixtures of endogenously-occurring metabolites, without added persistent radicals. In previous work using similar experimental conditions, UV-irradiated metabolite mixtures containing $_{13}$C-labeled PA and BA were hyperpolarized to significantly lower levels (i.e. 3.3 $\pm$ 0.5% − 5.2 $\pm$ 0.5% for $_{13}$C BA)[21]. The latter was a consequence of a less optimized sample preparation protocol, yielding a third of the radical concentration measured in this study, and the absence of frequency microwave modulation[21]. Indeed, we herein used a broad-band UV-light source delivering 40 times more power than the previous one and addition of glycerol-water to improve the DNP sample matrix. Although performed under different experimental conditions, previous studies using persistent radicals or non-endogenous nonpersistent radicals reported polarization levels between 22.2 $\pm$ 2.1%[38] and 30.1 $\pm$ 1.8%[25] for glucose and 7 $\pm$ 2%[33] to 28 $\pm$ 4%[35] for butyric acid. The current results are within a similar range and are thus promising. Although a dedicated optimization in terms of polarization level was not performed in the current study, polarization levels may be further increased by optimization of the sample formulation (i.e. components, concentrations and bead size), or by a detailed investigation of the microwave frequency modulation. However, previously reported improvements of DNP performance may conflict with the UV performance and nonpersistent radical generation: For example, using dimethyl sulfoxide as the glassing agent increased the DNP performance by 18-fold when polarizing $_{13}$C-labeled BA[35]. However, UV-Vis and ESR measurements have shown the photosensitive nature of dimethyl sulfoxide and its unsuitability as a glassing agent for UV-induced nonpersistent radical formulations using a broad-band UV source[24].

Hyperpolarized BA has been used before as a probe to study short-chain fatty acid cardiac metabolism[6,21,33,35,36], with different results in terms of the amount of observed metabolites. In the current study, hyperpolarization of BA with $\alpha$kV and $\alpha$kB reached sufficient SNR to observe cardiac metabolism with a similar metabolic profile observed in previous studies using persistent radicals at high magnetic field[6,33]. High magnetic field allows for a better spectral resolution and thus may facilitate the resolved detection of metabolites such as $_{13}$C labeled citrate and glutamate. However, $B_0$ inhomogeneities also increase which may complicate shimming procedures, especially around the heart. In the current study, the citrate resonance could not be reliably detected due to the shoulder of the neighboring [1-$_{13}$C]$\alpha$kB-hydrate resonance. Conversely, the absence of [1-$_{13}$C]$\alpha$kV-hydrate (Figure 6) enabled citrate detection when using $\alpha$kV. Note that the relative amount of $\alpha$kV in the BA sample formulation was much smaller compared to the amount of $\alpha$kB, which contributed to increased signal intensities of natural abundance $\alpha$kB resonances compared with those of $\alpha$kV. This can be appreciated when observing the relative signal ratio of [1-$_{13}$C]$\alpha$kV or [1-$_{13}$C]$\alpha$kB with [1-$_{13}$C]acetylcarnitine (Figure 6). Furthermore, the detection of [1-$_{13}$C]$\alpha$kV-hydrate in the non-pH-neutralized liquid state experiments compared with its absence in the pH-neutralized *in vivo* experiment, while the signal intensity of the [1-$_{13}$C]$\alpha$kB-hydrate resonance increased *in vivo* compared with the non-pH-neutralized liquid state experiment, demonstrates the sensitivity of the proposed polarizing agents to pH and illustrates the importance of pH optimization.

The use of endogenous $\alpha$kV and $\alpha$kB may be a versatile alternative to PA for studying fatty acid metabolism or metabolic processes where pyruvate is a metabolic product. Based on the peak locations of the $_{13}$C resonances of $\alpha$kV and $\alpha$kB, the potential detection of pyruvate as metabolic product would not be disturbed, especially in organs where $B_0$ inhomogeneities as well as motion-induced artifacts are reduced. Moreover, the enzyme PDH has a reduced or even absent affinity for $\alpha$kB and $\alpha$kV, respectively[43]. This was partially confirmed in previous experiments using hyperpolarized $_{13}$C-labeled $\alpha$kB that showed a decreased $_{13}$C-labeling of bicarbonate compared with PA[34]. Determining the *in vivo* affinity of PDH for hyperpolarized $_{13}$C-labeled $\alpha$kV, and potentially elucidating any metabolic interference of the proposed polarizing agents remains to be established. Nevertheless, radical-free dissolution DNP via the use of endogenous nonpersistent radicals provides a benefit at low-cost and may increase the duration of the hyperpolarized state by avoiding one filtration step in clinical applications. In addition, UV-induced radicals of PA were reported to quench with increasing temperature[18] and recombine at a threshold of 190 K[17], providing an additional benefit of UV-induced radicals in $\alpha$kV and $\alpha$kB to produce transportable hyperpolarized $_{13}$C-labeled substrates.

**CONCLUSION**

We conclude that the endogenous $\alpha$-keto acids $\alpha$kV and $\alpha$kB can be used as efficient endogenous nonpersistent radical following irradiation with UV light, achieving similar or higher $_{13}$C polarization when compared with PA and are thus potential candidates for translational clinical hyperpolarized MRI, enabling high polarization without requiring radical filtration. Cardiac metabolism of $_{13}$C labeled butyrate hyperpolarized with αkV or αkB demonstrated label propagation in a wide range of metabolites, demonstrating their potential as endogenous polarizing agents for *in vivo* radical-free hyperpolarized MRI.

**ACKNOWLEDGEMENTS**

This study was supported by funding received from the Swiss National Science Foundation (grant PZ00P3_167871 to JB), the Emma Muschamp foundation (JB), and the Swiss Heart foundation (grant FF18054 to JB). AC received funding from the Swiss National Foundation under the SPARK grant agreement no. CRSK-2_190547. We also thank Thanh Lê Phong for his assistance with the microwave modulation implementation, and Analina Raquel Da Silva, Mario Lepore and Stefanita-Octavian Mitrea for their assistance with the *in vivo* experiments. This study was supported in part by the CIBM of the EPFL, UNIL, UNIGE, HUG and CHUV.

Zanella *et al.*                                                                 Radical-free hyperpolarized MRI

Zanella *et al.*                                                                 Radical-free hyperpolarized MRI40. Takado Y, Cheng T, Bastiaansen JA, et al. Hyperpolarized 13C Magnetic Resonance Spectroscopy Reveals the Rate-Limiting Role of the Blood–Brain Barrier in the Cerebral Uptake and Metabolism of l-Lactate in Vivo. *ACS chemical neuroscience*. 2018;9(11):2554–2562.

41. Hu S, Zhu M, Yoshihara HA, et al. In vivo measurement of normal rat intracellular pyruvate and lactate levels after injection of hyperpolarized [1-13C] alanine. *Magnetic resonance imaging*. 2011;29(8):1035–1040.

42. Wishart DS, Feunang YD, Marcu A, et al. HMDB 4.0: the human metabolome database for 2018. *Nucleic acids research*. 2018;46(D1):D608–D617.

43. Bremmer J. Pyruvate dehydrogenase, substrate specificity and product inhibition. *European journal of biochemistry*. 1969;8(4):535–540.

44. Cheng T, Capozzi A, Takado Y, Balzan R, Comment A. Over 35% liquid-state 13 C polarization obtained via dissolution dynamic nuclear polarization at 7 T and 1 K using ubiquitous nitroxyl radicals. *Physical Chemistry Chemical Physics*. 2013;15(48):20819–20822.

45. Comment A, van den Brandt B vd, Uffmann K, et al. Design and performance of a DNP prepolarizer coupled to a rodent MRI scanner. *Concepts in Magnetic Resonance Part B: Magnetic Resonance Engineering: An Educational Journal*. 2007;31(4):255–269.

46. Cheng T, Mishkovsky M, Bastiaansen JA, et al. Automated transfer and injection of hyperpolarized molecules with polarization measurement prior to in vivo NMR. *NMR in biomedicine*. 2013;26(11):1582–1588.

47. Shaka AJ, Keeler J, Frenkiel T, Freeman RAY. An improved sequence for broadband decoupling: WALTZ-16. *Journal of Magnetic Resonance (1969)*. 1983;52(2):335–338.

48. Staewen RS, Johnson AJ, Ross BD, Parrish T, Merkle H, Garwood M. 3-D FLASH imaging using a single surface coil and a new adiabatic pulse, BIR-4. *Investigative radiology*. 1990;25(5):559–567.

49. Adeva B, Arik E, Ahmad S, et al. Large enhancement of deuteron polarization with frequency modulated microwaves. *Nuclear Instruments and Methods in Physics Research Section A: Accelerators, Spectrometers, Detectors and Associated Equipment*. 1996;372(3):339-343. doi:10.1016/0168-9002(95)01376-8

50. Hovav Y, Feintuch A, Vega S, Goldfarb D. Dynamic nuclear polarization using frequency modulation at 3.34 T. *Journal of Magnetic Resonance*. 2014;238:94–105.

51. Bornet A, Milani J, Vuichoud B, Linde AJP, Bodenhausen G, Jannin S. Microwave frequency modulation to enhance dissolution dynamic nuclear polarization. *Chemical Physics Letters*. 2014;602:63–67.

52. Ardenkjær-Larsen JH, Bowen S, Petersen JR, et al. Cryogen-free dissolution dynamic nuclear polarization polarizer operating at 3.35 T, 6.70 T, and 10.1 T. *Magnetic resonance in medicine*. 2019;81(3):2184–2194.
23

**TABLES**

*Table 1*

UV-induced radical build-up times and maximum nonpersistent radical concentrations obtained after 200 s of UV-irradiation in formulations containing 5 M precursor dissolved in glycerol-water. Data was acquired at 77 K using X-band ESR. Build-up rates were calculated from fitting a mono-exponential function to the radical concentration build-up curves (Figure 1e). Values represent mean and standard deviation, *n* = 3.

|  | **Build-up time (s)** | **Maximal concentration (mM)** |
|---|---|---|
| *α*kV | 30.9 ± 5.1 | 41.6 ± 0.6 |
| *α*kB | 37.0 ± 5.2 | 56.1 ± 2.7 |
| PA | 46.5 ± 1.4 | 55.0 ± 1.9 |

*Table 2*

Microwave center frequencies at which DNP maxima and minima occur at 7 T. Microwave frequency sweeps (Figure 3) were conducted on [U-$_{13}$C, U-$_2$H]Glc in glycerol-water samples. The microwave frequency was swept using either mono-chromatic microwave irradiation or frequency modulation with 40 MHz modulation amplitude at a 5 kHz rate.

|  | **Microwave frequency modulation** | **Positive DNP Maximum (GHz)** | **Negative DNP Maximum (GHz)** |
|---|---|---|---|
| **$\alpha$kV** | no | 196.73 | 196.89 |
|  | yes | 196.69 | 196.89 |
| **$\alpha$kB** | no | 196.69 | 196.89 |
|  | yes | 196.96 | 196.89 |
| **PA** | no | 196.67 | 196.87 |
|  | yes | 196.65 | 196.89 |

*Table 3*

Polarization build-up time constants at 7 T, 1.05 ± 0.02 K and liquid state polarization levels of the $C_1$, $C_{2-5}$ and $C_6$ resonances of [U-$^{13}C_6$, U-$^2H_7$]Glc at 9.4 T, 20°C. Liquid state enhancement was calculated as a ratio of hyperpolarized signal (rectangular RF excitation pulse of duration $\tau$ = 5 μs, RF excitation angle $\alpha$ = 5°) and thermal signal (64 averages of $\alpha$ = 90° with $\tau$ = 90 s, TR = 60 s). Values show mean and standard deviation over *n* = 3 data sets.

| *n* = 3 | Build-up time (s) | $C_1$ Glc liquid state polarization (%) | $C_{2-5}$ Glc liquid state polarization (%) | $C_6$ Glc liquid state polarization (%) |
|---|---|---|---|---|
| $\alpha$**kV** | 1.7k ± 0.5k | 9.4 ± 3.0 | 8.9 ± 2.1 | 8.0 ± 2.8 |
| $\alpha$**kB** | 1.3k ± 0.5k | 16.8 ± 0.4 | 16.3 ± 1.3 | 14.6 ± 1.7 |
| **PA** | 1.6k ± 0.5k | 13.9 ± 1.8 | 13.3 ± 1.5 | 11.5 ± 2.5 |
| **Conditions** | 7 T, 1.05 K | | 9.4 T, 293 K | |

*Table 4*

Solid state build-up times at 7T, 1.05 K are reported for [1-$^{13}$C]butyrate samples containing $\alpha$kV or $\alpha$kB as precursors. Room temperature liquid state enhancements, respectively liquid state polarization, were calculated after sample dissolution and transfer to a 9.4T MR scanner. Mean and average values are obtained from *n* data sets. Exemplary spectra for each sample are shown in Figure 5.

|  | Build-up time (s) | *n* | Liquid state enhancement | BA liquid state polarization (%) | *n* |
|---|---|---|---|---|---|
| *α*kV | 2.1k ± 0.3k | 4 | 14.6k ± 1.4k | 12.1 ± 1.1 | 3 |
| *α*kB | 3.3k ± 0.4k | 5 | 15.6k ± 1.7k | 12.9 ± 1.7 | 3 |
| **Conditions** | solid state: 7 T, 1.05 K | | liquid state: 9.4 T, 293 K | | |

*Table 5*

Chemical shifts of observed metabolites in the heart after injection of pH-neutralized hyperpolarized [1-$^{13}$C]butyrate *in vivo* using either $\alpha$kV or $\alpha$kB as polarizing agent (*in vivo* spectra displayed in Figure 6). [1-$^{13}$C]acetylcarnitine was used as reference peak (*) and assigned to 173.9 ppm. The resonance of natural abundance [1-$^{13}$C]$\alpha$kV-hydrate was not detected in these experiments (see also Supplemental Materials Figure S2).

| Metabolite | Chemical Shift (ppm) |
|---|---|
| natural abundance [2-$^{13}$C]$\alpha$kB | 209.3 |
| natural abundance [2-$^{13}$C]$\alpha$kV | 208.7 |
| [1-$^{13}$C]butyrate | 184.8 |
| [5-$^{13}$C]glutamate | 182.4 |
| [5-$^{13}$C]citrate | 179.8 |
| natural abundance [1-$^{13}$C]$\alpha$kB-hydrate | 178.1 |
| [1-$^{13}$C]butyrylcarnitine | 176.4 |
| [1-$^{13}$C]acetoacetate | 176.0 |
| [1-$^{13}$C]acetylcarnitine | 173.9* |
| natural abundance [1-$^{13}$C]$\alpha$kB | 172.1 |
| natural abundance [1-$^{13}$C]$\alpha$kV | 172.0 |

**FIGURES**

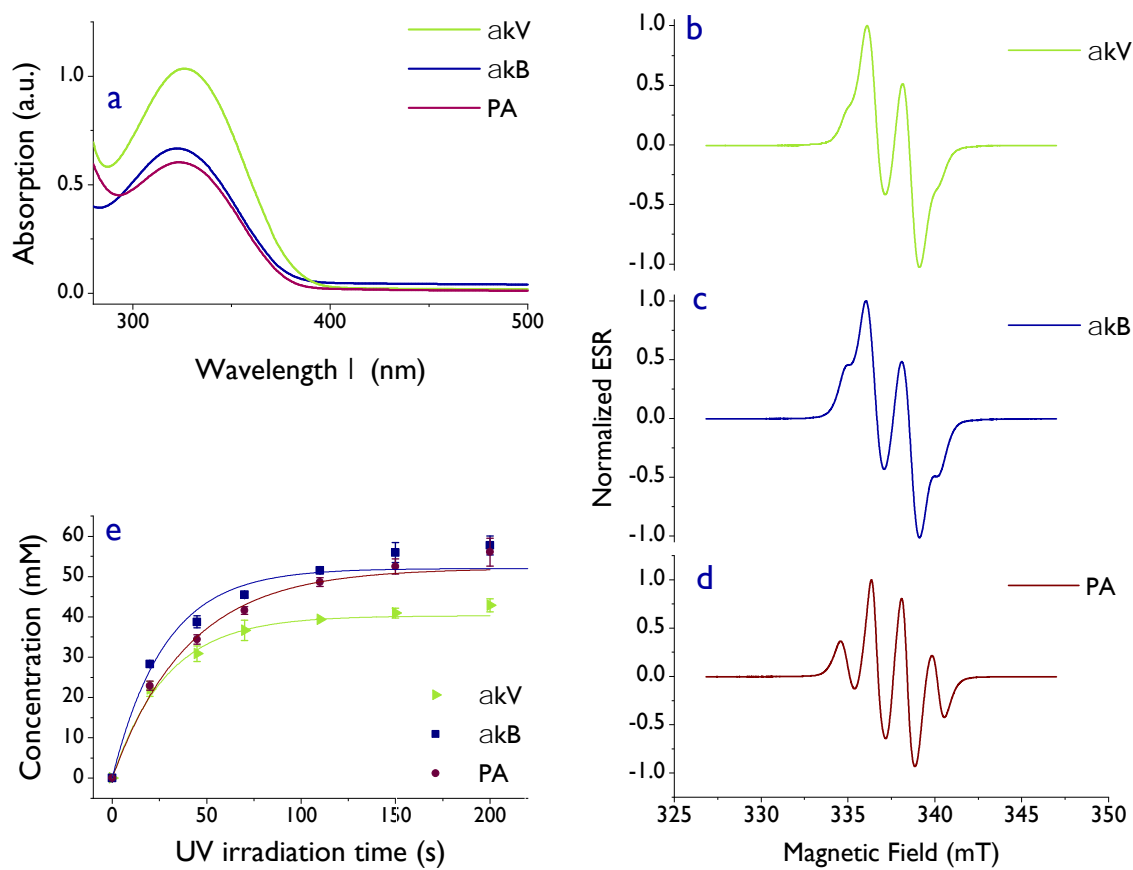

*Figure 1*

UV-Vis absorption spectra at room temperature and X-band ESR at 77 K. **a)** UV-Vis absorption spectra of 100 mM of radical precursor in glycerol-water using a 1 mm light path showing UV-light absorbance of $\alpha$kV, $\alpha$kB and PA. **b) c) d)** ESR spectra of the endogenous metabolites $\alpha$kV, $\alpha$kB and PA at 77 K after 200 s of UV irradiation with a 40 Wcm$_{-2}$ power UV-light source. **e)** Radical concentration build-up curves of 5 M precursor in glycerol-water upon UV-irradiation. Table 1 contains corresponding build-up times and maximum radical concentrations.

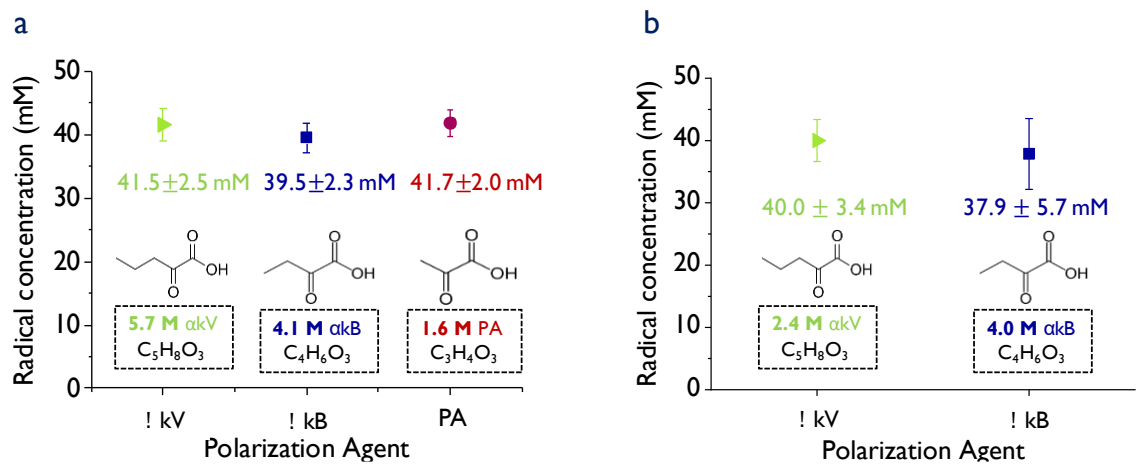

*Figure 2*

Sample formulations were optimized to obtain ~40mM of nonpersistent radicals after 200 s UV irradiation in **a)** [U-$_{13}$C, U-$_{2}$H]Glc and **b)** [1-$_{13}$C]BA. Underlying chemical structures are illustrated for each nonpersistent radical precursor, with the required concentrations indicated.

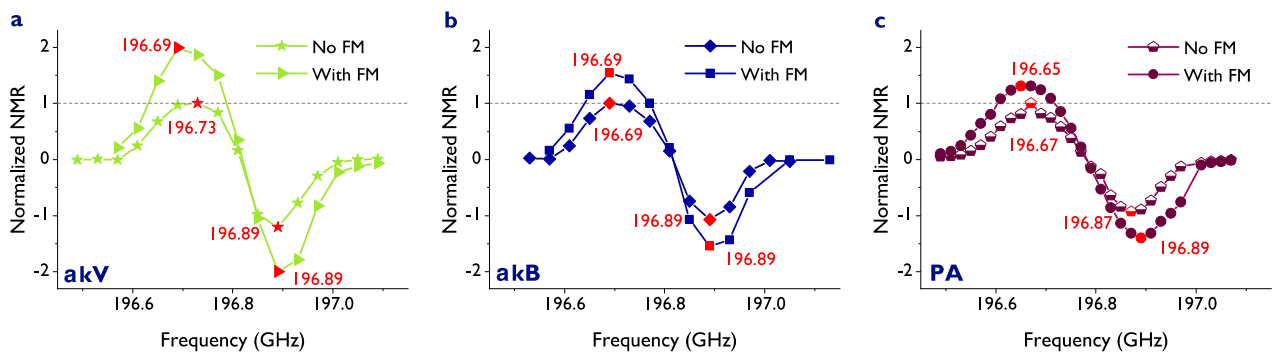

*Figure 3*

Hyperpolarized $_{13}$C signal as a function of microwave frequency with and without the application of frequency modulation (FM). Formulations containing [U-$_{13}$C, U-$_2$H]Glc in glycerol-water were hyperpolarized at 7 T and 4.2 K. FM was set to 40 MHz modulation amplitude at a frequency of 5 kHz. Microwave frequencies corresponding to observed DNP maxima and minima are reported (numbers in red) in Table 2. Hyperpolarization was achieved using the UV radicals **a)** $\alpha$kV, **b)** $\alpha$kB and **c)** PA.

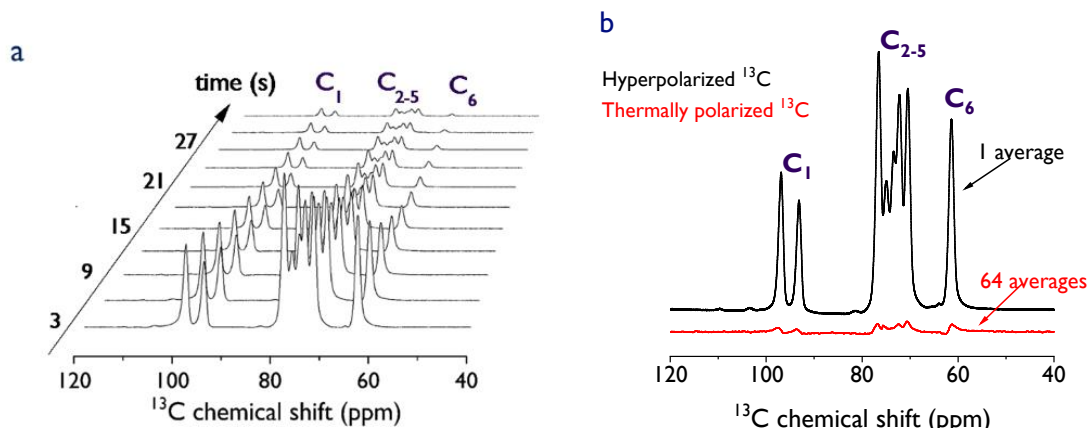

*Figure 4*

Hyperpolarized 13C MRS of [U-13C, U-2H]Glc sample formulations. **a)** Liquid state 13C signal evolution of [U-13C, U-2H]Glc hyperpolarized using $\alpha$kB as polarizing agent. **b)** Hyperpolarized 13C MR spectrum of [U-13C, U-2H]Glc acquired 3 s after dissolution (top) and the thermally polarized 13C spectrum (bottom). Acquisitions were performed at 9.4 T and $T = 20°C$

Zanella *et al.*                                                                 Radical-free hyperpolarized MRI

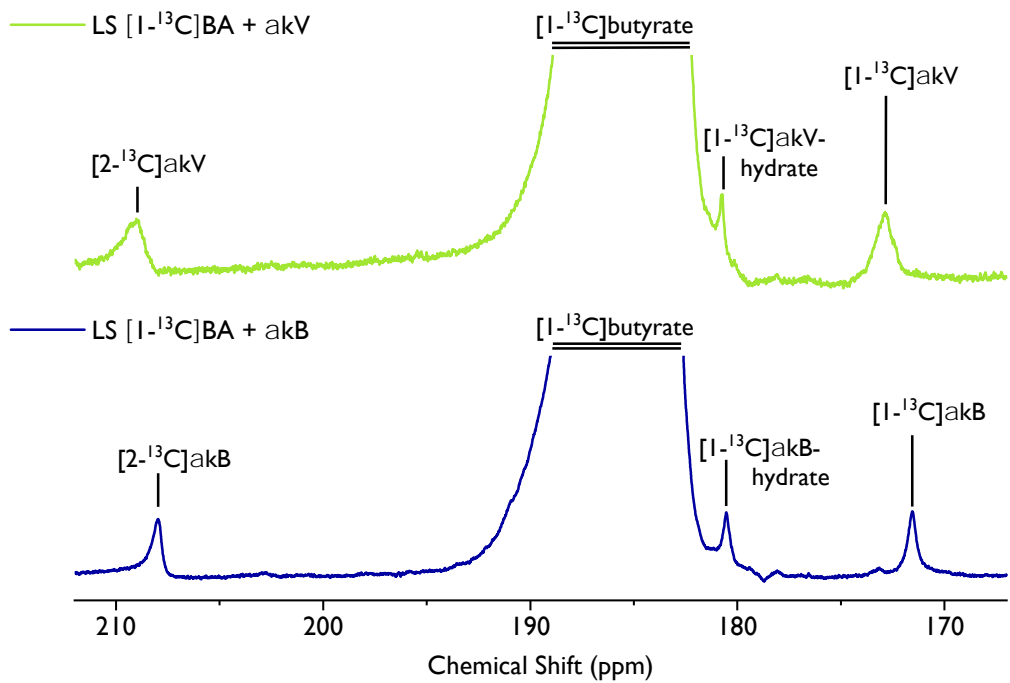

*Figure 5*

Liquid state $^{13}$C MR spectra of [1-$^{13}$C]butyrate hyperpolarized using $\alpha$kV (**top**) and $\alpha$kB (**bottom**) displaying the $^{13}$C-labeled substrate and precursor resonances. Spectra were acquired 3 s post dissolution and were line-broadened with 2 Hz. Solutions were acidic because pH neutralization was not performed in these experiments.

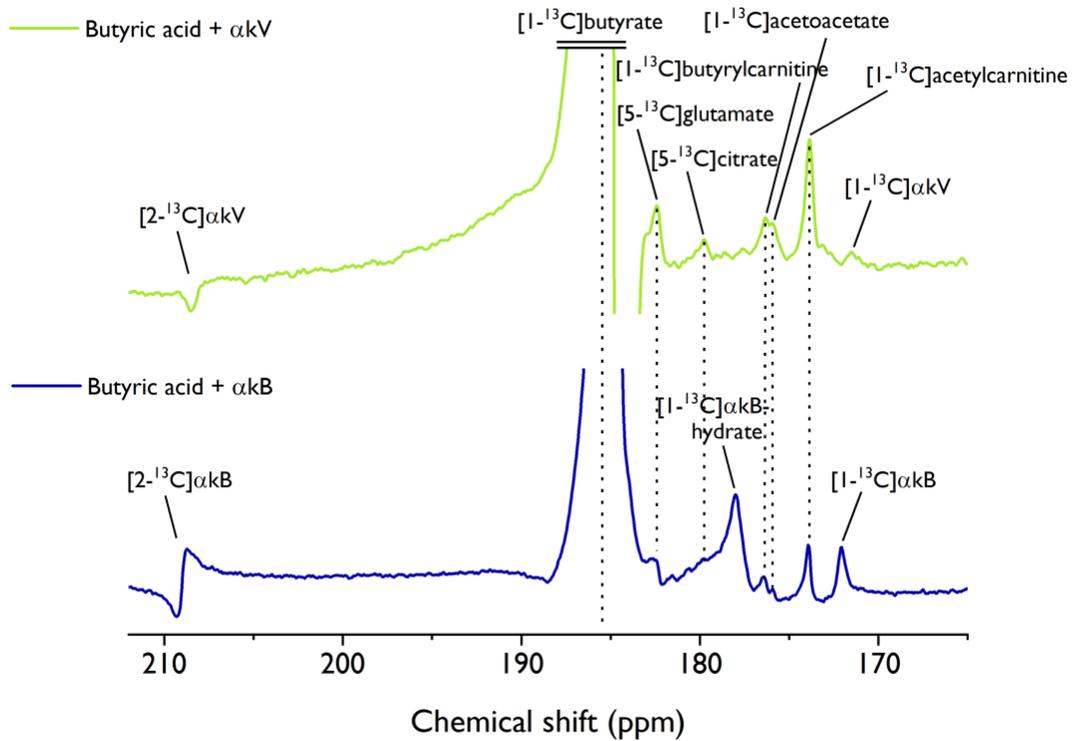

*Figure 6*

*In vivo* spectra of cardiac metabolism in two male Wistar rats following the injection of a radical-free hyperpolarized [1-13C]butyrate solution. Hyperpolarization via DNP was performed at 7 T, 1.05 K using the UV-induced nonpersistent radical $\alpha$kV (**top**) and $\alpha$kB (**bottom**). Spectra acquired at 9.4 T where [1-13C]acetylcarnitine resonances were visible were summed and line-broadened by a factor 20. Therefore, spectra acquired between 24 – 51 seconds (αkV) and 18 – 51 seconds post-dissolution (αkB) summed. Cardiac metabolism resulted in the detection of [1-13C]acetylcarnitine, [1-13C]acetoacetate, [1-13C]butyrylcarnitine and [5-13C]glutamate. [5-13C]citrate could not be detected in the αkB experiment, due to the proximity and phase of the αkB-hydrate resonance. The natural abundance resonances of the injected precursors αkB and αkV were also identified.

**SUPPLEMENTARY METHODS**

*Empirical procedure for optimizing sample formulation*

The sample formulation was empirically optimized to obtain 40 mM of radical concentration in beads, which has previously shown to be well-suited for nitroxyl radicals[1,2], with simultaneous satisfactory bead consistency. It is important to point out that samples were not specifically optimized to maximize radical yield, to minimize precursor concentration, to minimize the UV-irradiation time or to maximize achievable polarization levels.

In a first step, the maximum unlabeled Glc or BA concentration in glycerol-water was determined such that the solution created glassy beads upon freezing droplets of it in liquid nitrogen (according to a visual assessment). This solution acted as starting point for the next step.

In the second step, increasing volumes of radical precursors were added to reach our target value of 40 mM while ensuring that a maximum of 10% of the irradiated beads pulverized during UV-irradiation. If beads would not match the previous condition, glycerol-water was added in an iterative process until one possible mix yielding robust beads of 40 mM radical concentration was found. The UV-irradiation time of 200 s (for 6 ul beads) was chosen to avoid bulk pulverization (observed for 320 s of UV-irradiation) and to achieve the range of plateauing radical concentration (above 180 s UV-irradiation.

**SUPPLEMENTARY FIGURES**

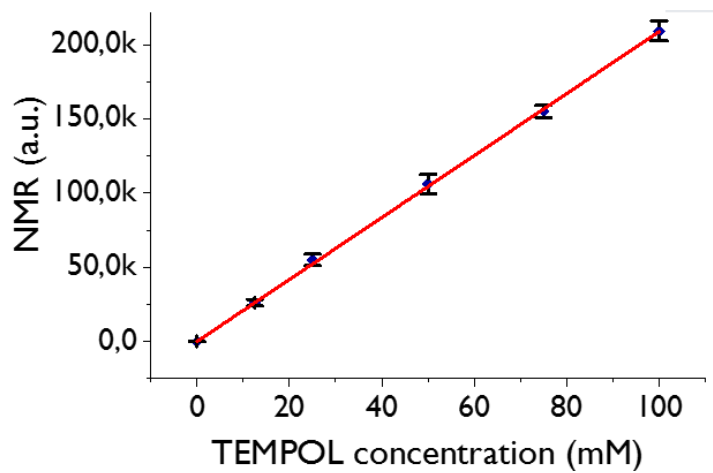

*Figure S1*

Concentration calibration curve of TEMPOL in glycerol-water (v/v 1:1) and linear fit ($R_2$ = 0.999, red) for n = 4 data sets. Data was acquired with X-band ESR at 77 K and subsequently used to quantify concentrations of UV-induced radicals.

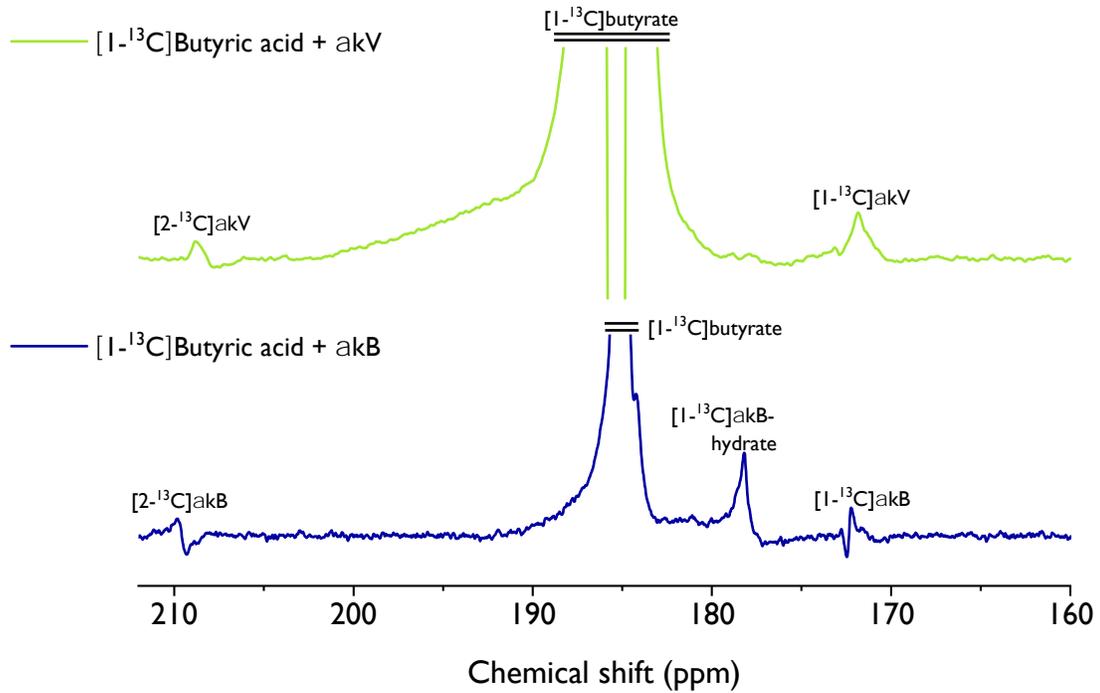

*Figure S2*

*In vivo* spectra acquired immediately after injection depict the injected $_{13}$C labeled substrate and the natural abundance $_{13}$C resonances of the polarizing agent $\alpha$kV (top) and $\alpha$kB (bottom). Metabolic products are absent in these spectra. Visible resonances are [1-$_{13}$C]butyrate, [1-$_{13}$C]$\alpha$kV, [2-$_{13}$C]$\alpha$kV, [1-$_{13}$C]$\alpha$kB-hydrate, [1-$_{13}$C]$\alpha$kB and [2-$_{13}$C]$\alpha$kB. The resonance of [1-$_{13}$C]$\alpha$kV-hydrate could not be detected.